\documentclass[11pt,a4paper]{article}
\usepackage{amssymb}
\usepackage{graphicx}

\input{tcilatex}
\begin{document}

\title{Symmetries, Integrability and Exact Solutions for Nonlinear Systems}
\author{R. Cimpoiasu, R. Constantinescu \\
University of Craiova, 13 A.I.Cuza, 200585 Craiova, Romania}
\date{}
\maketitle

\begin{abstract}
The paper intends to offer a general overview on what the concept of
integrability means for a nonlinear dynamical system and how the symmetry
method can be applied for approaching it. After a general part where key
problems as direct and indirect symmetry method or optimal system of
solutions are tackled out, in the second part of the lecture two concrete
models of nonlinear dynamical systems are effectively studied in order to
illustrate how the procedure is working out. The two models are the $2D$
Ricci flow model coming from the general relativity and the $2D$
convective-diffusion equation . Part of the results, especially concerning
the optimal systems of solutions, are new ones.

\end{abstract}

\section{Integrability and symmetries. Key aspects.}

\subsection{The concept of integrability for dynamical systems}

Dynamical systems described by nonlinear partial differential equations are
frequently used to model a wide variety of phenomena in physics, chemistry,
biology and other fields \cite{Scot}. The modelling process includes to find
solutions of those partial differential equations. If these solutions exist,
the diffential system is said to be integrable. Sometime it is difficult to
find a complete set of solutions and it is quite enough if one can decide on
the integrability of the system. There are many methods which can be used to
fulfill this aim: the Hirota's bilinear method, the Backlund transformation
method, the inverse scattering method, the Lax pair operator, the Painleve
analysis and others \cite{Her}. Each method has its own significant
properties. For example, while the Lax and the Painleve methods are moreover
testing the integrability, the Hirota's bilinear method is very efficient
for the effective determination of the multiple soliton solutions for a wide
class of nonlinear evolution equations \cite{Hiro}. As a conclusion, to
decide that a nonlinear differential equation is integrable, one of the
following situation shouls appear:

$(i)$ the existence of a number of functionally independent first
integrals/invariants equal to the order of the system in general and half
that for a Lagrangian system as a consequence of Liouville's Theorem;

$(ii)$ the existence of a sufficient number of Lie symmetries to reduce the
partial differential equation to an ordinary differential equation;

$(iii)$ the possession of the Painlev\'{e} property \cite{New}.

In this lecture the first two criteria will be investigated.

\subsection{The symmetry method for solving dynamical systems}

Many natural phenomena are described by a system of nonlinear partial
differential equations (pdes) which is often difficult to be solved
analytically, as there is no a general theory for completely solving of the
nonlinear pdes. One of the most useful techniques for finding exact
solutions of the dynamical systems described by nonlinear pdes is \textit{%
the symmetry method}. On the one hand, one can consider symmetry reduction
of differential equations and thus obtain classes of exact solutions. On the
other hand, by definition, a symmetry transforms solutions into solutions,
and thus symmetries can be used to generate new solutions from known ones.

Initially the symmetry method for solving partial differential equations was
developed for what is currently known as the \textit{Lie (classical)
symmetry method (CSM). }We shall present now a short introduction to this
approach \cite{Olver}.

Let us consider a $n$-th order partial differential system:%
\begin{equation}
\Delta _{\nu }(x,u^{(n)}[x])=0  \label{1.1}
\end{equation}%
where $x\equiv \{x^{i},$ $i=\overline{1,p}\}\subset R^{p}$ represent the
independent variables, while $u\equiv \{u^{\alpha },\alpha =\overline{1,q}%
\}\subset R^{q}$ the dependent ones. The notation $u^{(n)}$designates the
set of variables which includes $u$ and the partial derivatives of $u$ up to 
$n$-th order.

The general infinitesimal symmetry operator has the form: 
\begin{equation}
U=\dsum\limits_{i=1}^{p}\xi ^{i}(x,u)\frac{\partial }{\partial x^{i}}%
+\dsum\limits_{\alpha =1}^{q}\phi _{\alpha }(x,u)\frac{\partial }{\partial
u^{\alpha }}  \label{1.2}
\end{equation}%
The $n$-th extension of (\ref{1.2}) is given by: 
\begin{equation}
U^{(n)}=U+\dsum\limits_{\alpha =1}^{q}\dsum\limits_{J}\phi _{\alpha
}^{J}(x,u^{(n)})\frac{\partial }{\partial u_{J}^{\alpha }}  \label{1.3}
\end{equation}%
where 
\begin{equation}
u_{J}^{\alpha }=\frac{\partial ^{m}u^{\alpha }}{\partial x^{j_{1}}\partial
x^{j_{2}}..\partial x^{j_{m}}}  \label{1.4}
\end{equation}%
Also, in (\ref{1.4}) the second summation refers to all the multi-indices $%
J=(j_{1},...j_{m}),$with $1\leq $ $j_{m}$ $\leq $ $\ p,1\leq $ $m$ $\leq $ \ 
$n$ .The coefficient functions $\phi _{\alpha }^{J}$ are given by the
following formula:%
\begin{equation}
\phi _{\alpha }^{J}(x^{i},u^{(n)})=\mathcal{D}_{J}[\phi _{\alpha
}-\dsum\limits_{i=1}^{p}\xi ^{i}u_{i}^{\alpha }]+\dsum\limits_{i=1}^{p}\xi
^{i}u_{J,i}^{\alpha },\text{ }\alpha =\overline{1,q}  \label{1.5}
\end{equation}%
in which 
\begin{equation}
u_{i}^{\alpha }=\frac{\partial u^{\alpha }}{\partial x^{i}},\text{ }i=%
\overline{1,p}  \label{1.6}
\end{equation}%
\begin{equation}
u_{J,i}^{\alpha }=\frac{\partial u_{J}^{\alpha }}{\partial x^{i}}=\frac{%
\partial ^{m+1}u^{\alpha }}{\partial x^{i}\partial x^{j_{1}}\partial
x^{j_{2}}..\partial x^{j_{m}}}  \label{1.7}
\end{equation}

\begin{equation}
\mathcal{D}_{J}=\mathcal{D}\text{$_{j_{1}}$}\mathcal{D}\text{$_{j_{2}}$}...%
\text{\QTR{cal}{D}}_{j_{m}}=\frac{d\text{ }^{m}}{%
dx^{j_{1}}dx^{j_{2}}..dx^{j_{m}}}  \label{1.8}
\end{equation}%
The Lie symmetries represent the set of all the infinitesimal
transformations which keep invariant the differential system. The invariance
condition is:%
\begin{equation}
U^{(n)}[\Delta ]\shortmid _{\Delta =0}=0  \label{1.9}
\end{equation}%
The characteristic equations associated to general symmetry generator (\ref%
{1.2}) have the form:%
\begin{equation}
\frac{dx^{1}}{\xi ^{1}}=...=\frac{dx^{p}}{\xi ^{p}}=\frac{du^{1}}{\phi _{1}}%
=...=\frac{du^{q}}{\phi _{q}}  \label{1.10}
\end{equation}%
\qquad \qquad By integrating the characteristic system of ordinary
differential equations (\ref{1.10}), the invariants $I_{r},$ $r=\overline{%
1,(p+q-1)}$ of the analyzed system can be found. They are identified with
the constants of integration. Following this way, the set of similarity
variables is found in terms of which the original evolutionary equation with 
$p$ independent variables and $q$ dependent ones can be reduced to a set of
differential equations with $(p+q-1)$ variables. These are the similarity
reduced equation which generate the similarity solution of the analyzed
model.

There have been \textit{several generalizations} of the Lie symmetry method
which include:

$1)$ the \textit{non-classical symmetry method} (\textit{NSM}) (also
referred to as\textit{\ the conditional method}) of Bluman and Cole \cite%
{Bluman},

$2)$ the \textit{direct method} of Clarkson and Kruskal \cite{Clark},

$3)$ the \textit{differential constraint approach} of Olver and Rosenau \cite%
{Roseneau}

$4)$ the \textit{generalized conditional symmetry} method due to Fokas, Liu
and Zhdanov \cite{Fok}.

The basic idea of \textit{the nonclassical method} is that (\ref{2.1})
should be augmented with the invariance surface condition:%
\begin{equation}
Q^{\alpha }(x,u^{(1)})\equiv \phi _{\alpha }(x,u)-\dsum\limits_{i=1}^{p}\xi
^{i}(x,u)\frac{\partial u^{\alpha }}{\partial x^{i}}=0,\text{ }\alpha =%
\overline{1,q}  \label{1.101}
\end{equation}%
The $q-$tuple $Q=(Q^{1},Q^{2},...Q^{q})$\ is known as the characteristic of
the symmetry operator (\ref{1.2}). The invariance condition (\ref{1.9}) must
be applied taking into account that the constraints (\ref{1.101}) do exist.
The number of determining equations for the infinitesimals $\xi ^{i}(x,u),$\ 
$\phi _{\alpha }(x,u),$\ appearing in the nonclassical method is smaller
than for the classical method. The main difficulty of this approach is that
the determining equations are no longer linear. On the other hand, the NSM
may produce more solutions than the CSM, since any classical symmetry is a
nonclassical one, but not conversely.

The \textit{direct method }represents a direct, algorithmic, and nongroup
theoretic method for finding symmetry reductions. The relationship between
this direct method and the nonclassical method has been discussed in many
papers (e.g., \cite{Nucci}, \cite{Arri}). In particular, Levi and Winternitz 
\cite{Lw} established, using a group-theoretic explanation, that all new
solutions obtained by the direct method can also be obtained by the
nonclassical method. In fact, it has been shown in \cite{Pucci} that the
similarity solutions corresponding to the nonclassical groups should in
general constitute a larger family than that obtained by the direct method.

The \textit{differential constraint approach }proposed a generalization of
the nonclassical method. Its promoters shown that many known reduction
methods, including the classical and nonclassical methods, partial
invariance, and separation of variables can be placed into a general
framework. In their formulation, the original system of partial differential
equations can be enlarged by appending additional differential constraints
(side conditions), such that the resulting overdetermined system of partial
differential equations satisfying compatibility conditions.

As well, in further efforts to find new symmetries of PDEs which would lead
to additional new invariant solutions, much work has been done in the area
of higher-order symmetries. In particular, for an evolution equation in two
independent variables and one dependent variable has been introduced in \cite%
{Fok} the method of \textit{generalized conditional symmetries (GCS)} or
conditional Lie-B\"{a}cklund symmetries.

\subsection{Optimal system of solutions}

In general, when a differential equation admits a Lie group $\mathcal{G}_{r}$
and its Lie algebra $\mathcal{L}_{r}$ is of dimension $r$ $\rangle 1$, one
desires to minimize the search for invariant solutions by finding the
nonequivalent branches of solutions. This leads to the concept of \textit{%
optimal system}.

It is well known that for one-dimensional subalgebras, the problem of
finding an optimal system of subalgebras is essentially the same as the
problem of classifying the orbits of the adjoint transformations.

In Ovsiannikov \cite{Ovsian}, the \textit{global matrix of the adjoint
transformations} is used in constructing the one-dimensional optimal system.

In Olver \cite{Olver}, a slightly different technique is employed: it
consists in constructing a table, named the \textit{adjoint table}, showing
the separate adjoint actions of each element in $\mathcal{L}_{r}$ as it acts
on all the other elements .

The procedure reported in Ruggieri and Valenti \cite{RV}, is \ a mixed of
the above procedures and consists in constructing the \textit{global matrix
of the adjoint transformations} by means of the \textit{adjoint table}.

One of the advantages of the symmetry analysis is the possibility to find
solutions of the original pdes by solving odes. These odes, called \textit{%
reduced equations}, are obtained by introducing suitable new variables,
determined as invariant functions with respect to the infinitesimal
generators.

On the basis of the infinitesimal generators of the optimal systems of the
Lie algebras of analyzed model, we can construct the reduced odes of the
given model and find exact solutions.

\subsection{The inverse Lie symmetry problem}

Usually, \textit{the direct symmetry problem} of evolutionary equations is
considered for finding their exact solutions. It also known as the classical
symmetry method. Firstly, it consists in determining the Lie symmetry group
corresponding to a given evolutionary equation. Then, using the
characteristic equations could be obtained the Lie invariants associated to
each symmetry operator. Further these invariants, following the reduced
similarity procedure, determine the reduced equation which could be solve
and generates the similarity solution of the analyzed model.

Also, the \textit{inverse symmetry problem }\cite{noi}\textit{\ }could be
made. We ask the question: what is the largest class of evolutionary
equations which are equivalent from the point of view of their symmetries?.
So, this problem could be solved by imposing a concrete symmetry group to a
general analyzed model. With this condition, the general symmetry
determining equations could be solved and allow to determine all concrete
models which admit the same Lie symmetry group.

Let us consider a $2D$ dynamical system described by a second order partial
derivative equation of the general form:%
\begin{eqnarray}
u_{t}
&=&A(x,y,t,u)u_{xy}+B(x,y,t,u)u_{x}u_{y}+C(x,y,t,u)u_{2x}+D(x,y,t,u)u_{2y}+ 
\nonumber \\
&&+E(x,y,t,u)u_{y}+F(x,y,t,u)u_{x}+G(x,y,t,u)  \label{2.1}
\end{eqnarray}%
with $A(x,y,t,u),$ $B(x,y,t,u),$ $C(x,y,t,u),$ $D(x,y,t,u),$ $E(x,y,t,u),$ $%
F(x,y,t,u),$ $G(x,y,t,u)$ arbitrary functions of their arguments.

The general expression of the Lie symmetry operator which leaves (\ref{2.1})
invariant is:%
\begin{equation}
U(x,y,t,u)=\varphi (x,y,t,u)\frac{\partial }{\partial t}+\xi (x,y,t,u)\frac{%
\partial }{\partial x}+\eta (x,y,t,u)\frac{\partial }{\partial y}+\phi
(x,y,t,u)\frac{\partial }{\partial u}  \label{2.2}
\end{equation}%
Through loss the generality we can choose in the previous expression $%
\varphi \equiv 1.$Then, the generator (\ref{2.2}) becomes:%
\begin{equation}
U(x,y,t,u)=\frac{\partial }{\partial t}+\xi (x,y,t,u)\frac{\partial }{%
\partial x}+\eta (x,y,t,u)\frac{\partial }{\partial y}+\phi (x,y,t,u)\frac{%
\partial }{\partial u}  \label{2.3}
\end{equation}%
Following the symmetry theory \cite{Olver}, the second extension $U^{(2)}$
of (\ref{2.2}) has to be considered and the invariance condition of the
equation (\ref{2.1}) is given by the relation:%
\begin{eqnarray}
0
&=&U^{(2)}[u_{t}-A(x,y,t,u)u_{xy}-B(x,y,t,u)u_{x}u_{y}-C(x,y,t,u)u_{2x}-D(x,y,t,u)u_{2y}-
\nonumber \\
&&-E(x,y,t,u)u_{y}-F(x,y,t,u)u_{x}-G(x,y,t,u)]  \label{2.4}
\end{eqnarray}%
The previous relation has the equivalent expression:%
\begin{eqnarray}
0
&=&-A_{t}u_{xy}-B_{t}u_{x}u_{y}-C_{t}u_{2x}-D_{t}u_{2y}-E_{t}u_{y}-F_{t}u_{x}-G_{t}-A_{x}\xi u_{xy}-B_{x}\xi u_{x}u_{y}-
\nonumber \\
&&-C_{x}\xi u_{2x}-D_{x}\xi u_{2y}-E_{x}\xi u_{y}-F_{x}\xi u_{x}-G_{x}\xi
-A_{y}\eta u_{xy}-B_{y}\eta u_{x}u_{y}-C_{y}\eta u_{2x}-D_{y}\eta u_{2y}- 
\nonumber \\
&&-E_{y}\eta u_{y}-F_{y}\eta u_{x}-G_{y}\eta -A_{u}\phi u_{xy}-B_{u}\phi
u_{x}u_{y}-C_{u}\phi u_{2x}-D_{u}\phi u_{2y}-E_{u}\phi u_{y}-F_{u}\phi u_{x}-
\nonumber \\
&&-G_{u}\phi +\phi ^{t}-A\phi ^{xy}-C\phi ^{2x}-D\phi ^{2y}-B\phi
^{x}u_{y}-F\phi ^{x}-B\phi ^{y}u_{x}-E\phi ^{y}  \label{2.5}
\end{eqnarray}%
The functions $\phi ^{t},\phi ^{x},\phi ^{y},\phi ^{2x},\phi ^{2y},\phi
^{xy} $ will be determined using the general formulas:%
\begin{eqnarray}
\phi ^{t} &=&\mathcal{D}_{t}[\phi -u_{t}-\xi u_{x}-\eta u_{y}]+u_{2t}+\xi
u_{xt}+\eta u_{yt}  \nonumber \\
\phi ^{x} &=&\mathcal{D}_{x}[\phi -u_{t}-\xi u_{x}-\eta u_{y}]+u_{tx}+\xi
u_{2x}+\eta u_{xy}  \nonumber \\
\phi ^{y} &=&\mathcal{D}_{y}[\phi -u_{t}-\xi u_{x}-\eta u_{y}]+u_{ty}+\xi
u_{xy}+\eta u_{2y}  \nonumber \\
\phi ^{xy} &=&\mathcal{D}_{xy}[\phi -u_{t}-\xi u_{x}-\eta u_{y}]+u_{txy}+\xi
u_{xxy}+\eta u_{xyy}  \label{2.6} \\
\phi ^{2x} &=&\mathcal{D}_{2x}[\phi -u_{t}-\xi u_{x}-\eta u_{y}]+u_{txx}+\xi
u_{xxx}+\eta u_{xxy}  \nonumber \\
\phi ^{2y} &=&\mathcal{D}_{2y}[\phi -u_{t}-\xi u_{x}-\eta u_{y}]+u_{tyy}+\xi
u_{xyy}+\eta u_{yyy}  \nonumber
\end{eqnarray}%
By extending the relations (\ref{2.6}), substituting them into the condition
(\ref{2.5}) and then equating with zero the coefficient functions of various
monomials in derivatives of $u$, the following partial differential system
with$11$ equations is obtained:%
\begin{eqnarray*}
0 &=&\xi _{u} \\
0 &=&\eta _{u} \\
0 &=&B\eta _{x}-D\phi _{2u} \\
0 &=&B\xi _{y}-C\phi _{2u}
\end{eqnarray*}%
\begin{eqnarray*}
0 &=&A\eta _{y}-\eta A_{y}-A_{u}\phi +A\xi _{x}-\xi A_{x}++2D\xi _{y}+2C\eta
_{x}-A_{t} \\
0 &=&A\eta _{x}+2D\eta _{y}-\eta D_{y}-\xi D_{x}-D_{u}\phi -D_{t} \\
0 &=&-A\phi _{2u}+B\xi _{x}-B\phi _{u}+B\eta _{y}-B_{t}-B_{x}\xi -B_{u}\phi
-B_{y}\eta
\end{eqnarray*}%
\begin{eqnarray}
0 &=&-\eta _{t}+F\eta _{x}-B\phi _{x}+E\eta _{y}-E_{t}-E_{x}\xi -E_{y}\eta
-E_{u}\phi  \label{2.7} \\
&&+A\eta _{xy}-A\phi _{xu}+C\eta _{2x}+D\eta _{2y}-2D\phi _{yu}  \nonumber
\end{eqnarray}%
\begin{eqnarray*}
0 &=&-\xi _{t}-B\phi _{y}+F\xi _{x}+E\xi _{y}-F_{t}-F_{x}\xi -F_{y}\eta
-F_{u}\phi \\
&&A\xi _{xy}-A\phi _{yu}+C\xi _{2x}+D\xi _{2y}-2C\phi _{xu}
\end{eqnarray*}%
\begin{eqnarray*}
0 &=&\phi _{t}+G\phi _{u}-F\phi _{x}-E\phi _{y}-G_{t}-G_{x}\xi -G_{y}\eta
-G_{u}\phi \\
&&-A\phi _{xy}-C\phi _{2x}-D\phi _{2y}
\end{eqnarray*}%
The number of equations and of unknown functions which appear in the system (%
\ref{2.7}) is relatively high. Two approaches are now possible: $(i)$ to
find the symmetries of a given evolutionary equation, which means to choose
concrete forms for $A(x,y,t,u),$ $B(x,y,t,u),$ $C(x,y,t,u),$ $D(x,y,t,u),$ $%
E(x,y,t,u),$ $F(x,y,t,u),$ $G(x,y,t,u)$ and to use the system (\ref{2.7}) in
order to find the coefficient functions$~\xi (x,y,t),$ $\ \eta (x,y,t)$ and $%
\phi (x,y,t,u)$ of the Lie operator; $(ii)$ to solve the system (\ref{2.7})
taking as unknown variables\textbf{\ }$A(x,y,t,u),$ $B(x,y,t,u),$ $%
C(x,y,t,u),$ $D(x,y,t,u),$ $E(x,y,t,u),$ $F(x,y,t,u),$ $G(x,y,t,u)$ and
imposing a concrete form of the symmetry group. The first approach
represents \textit{the direct symmetry problem} and it is the usual one
followed in the study of the Lie symmetries of a given dynamical system. The
second approach, $(ii)$, represents the \textit{inverse symmetry problem}
and it is more special, allowing us to determine all the equations which are
equivalent from the point of view of the symmetry group they do admit.

\section{Applications}

In the next considerations, we will solve the direct and inverse Lie
symmetry problems for two $2D$ nonlinear models: the Ricci flow model and
the convective-diffusion equation.

\subsection{The Lie symmetry problems for 2D Ricci flow model}

One of the most fruitful models used in study of the black holes and in the
attempt of obtaining a quantum theory of gravity is connected with the 
\textit{Ricci flow equations }\cite{Bakas1}.

We will investigate a $2D$ model for the Ricci flow equation, a nonlinear
parabolic equation obtained when the components of the metric tensor $%
g_{\alpha \beta }$ are deformed following the equation: 
\begin{equation}
\frac{\partial }{\partial t}g_{\alpha \beta }=-R_{\alpha \beta }  \label{3.1}
\end{equation}%
where $R_{\alpha \beta }$ is the Ricci tensor for the $n$-dimensional
Riemann space. The metric tensor of the space $g_{\alpha \beta }$ will be
connected with the Riemann metric in the conformal gauge: 
\begin{equation}
ds^{2}=g_{\alpha \beta }dx^{\alpha }dx^{\beta }=\frac{1}{2}\exp \{\Phi
(X,Y,t)\}(dX^{2}+dY^{2})  \label{3.2}
\end{equation}%
The \textquotedblright potential\textquotedblright\ $\Phi (X,Y,t)$ satisfies
the equation: 
\begin{equation}
\frac{\partial }{\partial t}e^{\Phi }=\triangle \Phi  \label{3.3}
\end{equation}%
It has been noticed \cite{Bakas} that the equation (\ref{3.3}) is pretty
similar with the Toda equation describing the integrable interaction of a
collection of two dimensional fields $\{\Phi _{i},i=1,2\}$ coupled by a
Cartan matrix $(K_{ij})$: 
\begin{equation}
\sum_{j}K_{ij}e^{\Phi _{j}(X,Y)}=\triangle \Phi _{i}(X,Y)  \label{3.4}
\end{equation}%
Introducing the field $u(x,y,t)$ given by

\begin{equation}
u(x,y,t)=e^{\Phi }  \label{3.5}
\end{equation}%
the equation (\ref{3.3}) takes the form: 
\begin{equation}
u_{t}=(\ln u)_{xy}  \label{3.6}
\end{equation}%
An equivalent form for the previous equation, which will be used in the next
considerations of the paper, is: 
\begin{equation}
u_{t}=\frac{u_{xy}}{u}-\frac{u_{x}u_{y}}{u^{2}}  \label{3.7}
\end{equation}%
The previous equation could be derived from the general one (\ref{2.1}) by
choosing the following particular coefficient functions:%
\begin{eqnarray}
A(x,y,t,u) &=&\frac{1}{u},B(x,y,t,u)=-\frac{1}{u^{2}},  \nonumber \\
C(x,y,t,u) &=&D(x,y,t,u)=E(x,y,t,u)=F(x,y,t,u)=G(x,y,t,u)\equiv 0
\label{3.8}
\end{eqnarray}

\subsubsection{Lie symmetries for 2D Ricci flow model}

To finding the \textit{Lie symmetry operators} for the Ricci flow model (\ref%
{3.7}) we have to solved the general determining system (\ref{2.7}) in the
conditions (\ref{3.8}). The solution is represented by the coefficient
functions $\xi (x,y,t),$ $\eta (x,y,t),$ $\phi (x,y,t,u)$ which determine
the Lie symmetry operator (\ref{2.2}). It has the form:%
\begin{equation}
U=\frac{\partial }{\partial t}+\xi (x)\frac{\partial }{\partial x}+\eta (y)%
\frac{\partial }{\partial y}-u[\xi _{x}(x)+\eta _{y}(y)]\frac{\partial }{%
\partial u}  \label{3.9}
\end{equation}

As $U$ contains coefficients in the form of two arbitrary functions $\{\xi
,\eta \}$, we deal with an infinite number of symmetry operators. The action
of $U$ can be split in various \textquotedblright sectors\textquotedblright
, depending on the concrete form we might choose for these functions.

Let us consider the \textit{linear sector} of the Lie symmetries in which
the forms of the coefficient functions of the symmetry generator (\ref{2.2})
are:%
\begin{equation}
\varphi =1,\text{ }\xi =mx+c_{1},\text{ }\eta =vy+c_{2},\text{ }\phi =-(m+v)u
\label{3.10}
\end{equation}%
with $m,v,k,c_{1},c_{2},c_{3}$ arbitrary constants.

The general Lie symmetry operator (\ref{2.2}) becomes:%
\begin{equation}
U(x,y,t,u)=\frac{\partial }{\partial t}+(mx+c_{1})\frac{\partial }{\partial x%
}+(vy+c_{2})\frac{\partial }{\partial y}-(m+v)u\frac{\partial }{\partial u}
\label{3.11}
\end{equation}%
Consequently, the nonlinear Ricci flow equation (\ref{3.7}) admits the $4-$%
dimensional Lie algebra spanned by the independent operators shown below:%
\begin{equation}
\,V_{1}=x\frac{\partial }{\partial x}-u\frac{\partial }{\partial u},V_{2}=%
\frac{\partial }{\partial x},\,V_{3}=y\frac{\partial }{\partial y}-u\frac{%
\partial }{\partial u},V_{4}=\frac{\partial }{\partial y}  \label{3.12}
\end{equation}%
The forms of the operators $V_{i},$ $i=\overline{1,4}$ suggest their
significations: $V_{2},V_{4}$ generate the symmetry of space translations, $%
V_{1},U_{3}$ are associated with the scaling transformations.

When the Lie algebra of these operators is computed, the only non-vanishing
relations are: 
\begin{equation}
\,\,[V_{2},V_{1}]=V_{2},[V_{4},V_{3}]=V_{4}  \label{3.13}
\end{equation}

\subsubsection{Optimal system of subalgebras for 2D Ricci flow model}

It is well known that reduction of he independent variables by one is
possible using any linear combination of the generators of symmetry (\ref%
{3.12}) $V_{i},i=\overline{1,4}$. We will construct a set of minimal
combinations known as optimal system \cite{Olver}. To construct the optimal
system we need the commutators of the admitted symmetries given in the Table
1.

\bigskip

\begin{tabular}{ccccc}
$\lbrack V_{i},V_{j}]$ & $V_{1}$ & $V_{2}$ & $V_{3}$ & $V_{4}$ \\ 
$V_{1}$ & $0$ & $-V_{2}$ & $0$ & $0$ \\ 
$V_{2}$ & $V_{2}$ & $0$ & $0$ & $0$ \\ 
$V_{3}$ & $0$ & $0$ & $0$ & $-V_{4}$ \\ 
$V_{4}$ & $0$ & $0$ & $V_{4}$ & $0$%
\end{tabular}
\ \ \ \ \ \ Table1: Lie brackets of the admitted symmetry algebra

An optimal system of a Lie algebra is a set of $l-$dimensional subalgebras
such that every $l-$dimensional subalgebra is equivalent to a unique element
of the set under some element of the adjoint representation. The adjoint
representation of a Lie algebra $\{V_{i},i=1,...,4\}$ is constructed using
the formula \cite{Olver}:%
\begin{equation}
Ad(\exp (\varepsilon V_{i}))V_{j}=\sum_{n}\frac{\varepsilon ^{n}}{n!}%
(adV_{i})^{n}V_{j}=V_{j}-\varepsilon \lbrack V_{i},V_{j}]+\frac{\varepsilon
^{2}}{2!}[V_{i},[V_{i},V_{j}]]-...  \label{5.1}
\end{equation}%
Let us consider the linear combination of the symmetry generators:%
\begin{equation}
V=a_{1}V_{1}+a_{2}V_{2}+a_{3}V_{3}+a_{4}V_{4}  \label{5.2}
\end{equation}%
Our task is to simplify as many of the coefficients $a_{i}$ as possible
through judicious applications of adjoint maps to $V.$Suppose first that $%
a_{1}\neq 0$ in (\ref{5.2}). One may re-scale $a_{1}$ such that $a_{1}=1.$
We start with the combination:%
\begin{equation}
V^{(1)}=V_{1}+a_{2}V_{2}+a_{3}V_{3}+a_{4}V_{4}  \label{5.3}
\end{equation}%
If we act on $V^{(1)}$ by $Ad(\exp (a_{2}V_{2})),$ we can make the
coefficient of $V_{2}$ vanish:%
\begin{equation}
V^{(2)}=V_{1}+a_{3}V_{3}+a_{4}V_{4}  \label{5.4}
\end{equation}%
Next, we act on $V^{(2)}$ by $Ad(\exp (\frac{a_{4}}{a_{3}}V_{4}))$ to cancel
the coefficient of $V_{4},$ leading to the operator:%
\begin{equation}
V^{(3)}=V_{1}+a_{3}V_{3}  \label{5.5}
\end{equation}%
Using the adjoint representation (\ref{5.1}), no further simplification is
possible. Consequently, the $1-$dimensional subalgebra spanned by $V$ with $%
a_{1}\neq 0$ is equivalent to the one spanned by $V_{1}+\beta V_{3},$ $\beta
\in R.$

The remaining $1-$dimensional subalgebras are spanned by operators with $%
a_{1}=0$ which have the expressions:%
\begin{equation}
V^{(4)}=a_{2}V_{2}+a_{3}V_{3}+a_{4}V_{4}  \label{5.6}
\end{equation}%
Let us assume that $a_{2}\neq 0$ and scale to make $a_{2}=1.$ Now we act on $%
V^{(4)}$ by $Ad(\exp (\frac{a_{4}}{a_{3}}V_{4}))$ so that it is equivalent
with the operator:%
\begin{equation}
V^{(5)}=V_{2}+a_{3}V_{3}  \label{5.7}
\end{equation}%
No further simplification is possible. Consequently, the $1-$dimensional
subalgebra spanned by $V$ with $a_{2}\neq 0$ is equivalent to the one
spanned by $V_{2}+\alpha V_{3},$ $\alpha \in R.$

If we consider the case $a_{1}=a_{2}=a_{3}=0,$ $a_{3}\neq 0,a_{3}=1$, the
following generator is obtained:%
\begin{equation}
V^{(6)}=V_{3}+a_{4}V_{4}  \label{5.8}
\end{equation}%
Acting on $V^{(6)}$ by $Ad(\exp (a_{4}V_{4})),$ we obtain the operator $%
V_{3} $ which represents the next subalgebra of the optimal system.

Finally, let us consider the last case $a_{1}=a_{2}=a_{3}=0,$ $a_{4}\neq
0,a_{4}=1$ in (\ref{5.2}). Results the last subalgebra $V_{4}.$

In conclusion, the optimal system of $1-$dimensional subalgebras has the
form:%
\begin{equation}
\{V_{2}+\alpha V_{3},V_{1}+\beta V_{3},V_{3},V_{4}\}  \label{5.9}
\end{equation}

\subsubsection{Invariant solutions for $2D$ Ricci flow}

Let us pass now to the problem of the\textit{\ invariant quantities. }\ We
shall analyze the invariants associated with the optimal system of symmetry
operators (\ref{5.9}).

\begin{itemize}
\item The operator $V_{2}+\alpha V_{3}$ from (\ref{5.9}) has the
characteristic equations:%
\begin{equation}
\frac{dt}{0}=\frac{dx}{1}=\frac{dy}{\alpha y}=\frac{du}{-\alpha u}
\label{5.10}
\end{equation}%
By integrating these equations result $3$ invariants with expressions:%
\begin{equation}
I_{1}=t,\text{ }I_{2}=ye^{-\alpha x},\text{ }I_{3}=yu  \label{5.11}
\end{equation}%
By introducing the similarity variable $z\equiv $ $I_{2}=ye^{-\alpha x}$,
designating the invariant $I_{3}$ $=h(t,z)$ as a function of the other ones,
the following solution is obtained:%
\begin{equation}
u(t,x,y)=\frac{h(t,z)}{y}  \label{5.12}
\end{equation}%
Setting the derivatives of (\ref{5.12}) into the Ricci equation (\ref{3.7}),
results the similarity reduced equation for $h(t,z)$ with the form:%
\begin{equation}
h_{t}h^{2}-\alpha z^{2}hh_{2z}-\alpha z^{2}h_{z}^{2}+\alpha zhh_{z}=0
\label{5.13}
\end{equation}%
The solution of the previous equation is:%
\begin{equation}
h(t,z)=-\frac{1}{2}\left( r_{3}t+\frac{r_{2}r_{3}}{2r_{1}}\right) \left(
-1+\tanh ^{2}\left( \frac{\sqrt{\alpha r_{3}}(r_{4}-\ln z)}{2\alpha }\right)
\right)  \label{5.15}
\end{equation}%
with $\alpha ,r_{1},r_{2},r_{3}$ arbitrary constants and $z$ the similarity
variable.
\end{itemize}

Consequently, the invariant solution corresponding to operator $V_{2}+\alpha
V_{3}$ has the final form:%
\begin{equation}
u(t,x,y)=-\frac{1}{2y}\left( r_{3}t+\frac{r_{2}r_{3}}{2r_{1}}\right) \left(
-1+\tanh ^{2}\left( \frac{\sqrt{\alpha r_{3}}(r_{4}-\alpha x+\ln y)}{2\alpha 
}\right) \right)  \label{5.16}
\end{equation}

\begin{itemize}
\item The operator $V_{1}+\beta V_{3}$ from (\ref{5.9}) has the
characteristic equations:%
\begin{equation}
\frac{dt}{0}=\frac{dx}{x}=\frac{dy}{\beta y}=\frac{du}{-(1+\beta )u}
\label{5.17}
\end{equation}%
In this second case, following the same procedure, we obtain also $3$
independent invariants with expressions:%
\begin{equation}
I_{1}=t,\text{ }I_{2}=yx^{-\beta },\text{ }I_{3}=y^{(1+\beta )/\beta }u
\label{5.18}
\end{equation}%
By introducing the similarity variable $z\equiv $ $I_{2}=yx^{-\beta }$,
designating the invariant $I_{3}$ $=h(t,z)$ as a function of the other ones,
the following solution is obtained:%
\begin{equation}
u(t,x,y)=h(t,z)y^{-(1+\beta )/\beta }  \label{5.19}
\end{equation}
\end{itemize}

Setting the derivatives of (\ref{5.19}) into the Ricci equation (\ref{3.7}),
results the following $(1+1)$ reduced equation for $h(t,z)$:%
\begin{equation}
h_{t}h^{2}z^{(-1/\beta -2)}+\beta hh_{2z}+\beta z^{-1}hh_{z}-\beta
h_{z}^{2}=0  \label{5.20}
\end{equation}%
The solution of the previous equation is:%
\begin{equation}
h(t,z)=-\frac{\left( p_{1}t+p_{2}\right) }{2p_{3}^{2}p_{1}\beta }z^{1/\beta
}\left( -1+\tanh ^{2}\left( \frac{p_{4}\beta -\ln (z)}{2p_{3}\beta }\right)
\right)  \label{5.21}
\end{equation}%
with $\beta ,p_{1},p_{2},p_{3},p_{4}$ arbitrary constants and $z$ the
similarity variable.

Consequently, by using (\ref{5.19}) the invariant solution corresponding to
the operator $V_{1}+\beta V_{3}$ has the final form:%
\begin{equation}
u(t,x,y)=-\frac{1}{2p_{3}^{2}p_{1}\beta }\frac{(p_{1}t+p_{2})}{xy}\left(
-1+\tanh ^{2}\left( \frac{p_{4}\beta -\ln y+\beta \ln (x)}{2p_{3}\beta }%
\right) \right)  \label{5.22}
\end{equation}

\begin{itemize}
\item Because (\ref{3.7}) is symmetric in $\ x$ and $y$, there is also a
second similarity solution of the form: 
\begin{equation}
u(x,y)=\frac{g_{3}(x)}{y},\text{ }\forall g_{3}(x)  \label{3.16}
\end{equation}

which is generated by the symmetry operator $V_{3}$ from (\ref{3.12})$.$

\item Again, by the reason of symmetry in $x$ and $y$ of the analyzed model (%
\ref{3.7}), the last similarity solution associated to the symmetry operator 
$V_{4}$ from (\ref{3.12}), is generated as below:%
\begin{equation}
u(x)=g_{4}(x),\text{ }\forall g_{4}(x)  \label{3.163}
\end{equation}
\end{itemize}

\subsubsection{The class of equations with Ricci type symmetries}

Now, our aim is to find the class of equations with generic form (\ref{2.1})
which admit Ricci type symmetries (\ref{3.10}). Consequently, we have to
solve the system (\ref{2.7}) taking as unknown functions\textbf{\ }$%
A(x,y,t,u),$ $B(x,y,t,u),$ $C(x,y,t,u),$ $D(x,y,t,u),$ $E(x,y,t,u),$ $%
F(x,y,t,u),$ $G(x,y,t,u)$ and imposing a concrete form of the symmetry group.

Let us consider the \textit{linear sector} for the Lie symmetries where
coefficient functions of symmetry operators have the general expressions:%
\begin{equation}
\varphi =1,\text{ }\xi =mx+c_{1},\text{ }\eta =vy+c_{2},\text{ }\phi
=ku+c_{3}  \label{3.17}
\end{equation}%
with $m,v,k,c_{1},c_{2},c_{3}$ arbitrary constants.

\textbf{Remark 1: }For the concrete values $k=-(m+v),c_{3}=0$ and
nonvanishing values for all the other constants $m,v,k,c_{1},c_{2},$ the
linearization (\ref{3.17}) is reduced to (\ref{3.10}) considered for the
Ricci flow model. The results will be particularized for this case when the
Ricci symmetries have been imposed.

With the choice (\ref{3.17}), the general differential system (\ref{2.7})
has for the "unknown" functions $A(x,y,t,u),$ $B(x,y,t,u),$ $C(x,y,t,u),$ $%
D(x,y,t,u),$ $E(x,y,t,u),$ $F(x,y,t,u),$ $G(x,y,t,u)$ solutions which can be
expressed in terms of $14$ arbitrary functions $\{\mathcal{F}_{j}^{\prime },$
$\mathcal{F}_{j}^{\prime \prime },j=1,...,7\}$. They are the general forms:%
\begin{eqnarray}
A(x,y,t,u) &=&\mathcal{F}_{1}^{\prime }\left( X^{\prime },Y^{\prime
},Z^{\prime }\right) \xi ^{\frac{m+v}{m}}+\mathcal{F}_{1}^{\prime \prime
}(X",Y",Z")\eta ^{\frac{m+v}{v}}  \nonumber \\
B(x,y,t,u) &=&\mathcal{F}_{2}^{\prime }\left( X^{\prime },Y^{\prime
},Z^{\prime }\right) \xi ^{\frac{m+v-k}{m}}+\mathcal{F}_{2}^{\prime \prime
}(X",Y",Z")\eta ^{\frac{m+v-k}{v}}  \nonumber \\
C(x,y,t,u) &=&\mathcal{F}_{3}^{\prime }\left( X^{\prime },Y^{\prime
},Z^{\prime }\right) \xi ^{2}+\mathcal{F}_{3}^{\prime \prime }(X",Y",Z")\eta
^{2}  \nonumber \\
D(x,y,t,u) &=&\mathcal{F}_{4}^{\prime }\left( X^{\prime },Y^{\prime
},Z^{\prime }\right) \xi ^{\frac{2v}{m}}+\mathcal{F}_{4}^{\prime \prime
}(X",Y",Z")\eta ^{\frac{2m}{v}}  \label{3.18} \\
E(x,y,t,u) &=&\mathcal{F}_{5}^{\prime }\left( X^{\prime },Y^{\prime
},Z^{\prime }\right) \xi ^{\frac{v}{m}}+\mathcal{F}_{5}^{\prime \prime
}(X",Y",Z")\eta ^{\frac{m}{v}}  \nonumber \\
F(x,y,t,u) &=&\mathcal{F}_{6}^{\prime }\left( X^{\prime },Y^{\prime
},Z^{\prime }\right) \xi ^{\frac{v}{m}}+\mathcal{F}_{6}^{\prime \prime
}(X",Y",Z")\eta ^{\frac{m}{v}}  \nonumber \\
G(x,y,t,u) &=&\mathcal{F}_{7}^{\prime }\left( X^{\prime },Y^{\prime
},Z^{\prime }\right) \xi ^{\frac{k}{m}}+\mathcal{F}_{7}^{\prime \prime
}(X",Y",Z")\eta ^{\frac{k}{v}}  \nonumber
\end{eqnarray}%
where 
\begin{eqnarray}
X^{\prime } &\equiv &\frac{\eta \xi ^{\left( -\frac{v}{m}\right) }}{v}%
,Y^{\prime }\equiv \frac{mt-\ln \xi }{m},Z^{\prime }\equiv \frac{\phi \xi
^{\left( -\frac{k}{m}\right) }}{k},  \nonumber \\
X" &\equiv &\frac{\xi \eta ^{\left( -\frac{m}{v}\right) }}{m},Y"\equiv \frac{%
vt-\ln \eta }{v},Z"\equiv \frac{\phi \eta ^{\left( -\frac{k}{v}\right) }}{k}.
\label{3.19}
\end{eqnarray}

\textbf{Remark 2: }In the case $k=0$, the arguments $Z^{\prime },Z"$ are
changing and they become: 
\begin{equation}
Z^{\prime }=\frac{mu-c_{3}\ln \xi }{m},Z"=\frac{vu-c_{3}\ln \eta }{v}.
\label{3.20}
\end{equation}

It is useful to choose particular solutions. Let us consider the cases:

$Case1:c_{3}=0$ and $(m+v)$ is a multiple of the constant $k$, that is: 
\begin{equation}
m+v=k\cdot n,\text{ }(\forall )n\in \mathbf{Z\backslash \{}0\mathbf{\}}
\label{3.21}
\end{equation}%
Following the general solution (\ref{3.18})-(\ref{3.19}), the concrete forms
for $A(x,y,t,u),$ $B(x,y,t,u),$ $C(x,y,t,u),$ $D(x,y,t,u),$ $E(x,y,t,u),$ $%
F(x,y,t,u),$ $G(x,y,t,u)$ may be in this case:%
\begin{eqnarray}
A(x,y,t,u) &=&\left[ \frac{1}{2}u^{n}\xi ^{(-\frac{n}{m}k)}\right] \xi ^{(%
\frac{n}{m}k)}+\left[ \frac{1}{2}u^{n}\eta ^{(-\frac{n}{v}k)}\right] \eta ^{(%
\frac{n}{v}k)}=u^{n}  \nonumber \\
B(x,y,t,u) &=&\left[ \frac{1}{2}nu^{(n-1)}\xi ^{(-\frac{n-1}{m}k)}\right]
\xi ^{(\frac{n-1}{m}k)}+\left[ \frac{1}{2}nu^{(n-1)}\eta ^{(-\frac{n-1}{v}k)}%
\right] \eta ^{(\frac{n-1}{v}k)}=nu^{(n-1)}  \label{3.22}
\end{eqnarray}%
\[
C(x,y,t,u)=D(x,y,t,u)=E(x,y,t,u)=F(x,y,t,u)=G(x,y,t,u)=0 
\]%
where the parameter $k$ has the form (\ref{3.21}).

The $2D$ general equation (\ref{2.1}) takes the particular form:%
\begin{equation}
u_{t}=u^{n}u_{xy}+nu^{(n-1)}u_{x}u_{y}=(u^{n}u_{x})_{y}  \label{3.23}
\end{equation}

\textbf{Remark 3}: For $n=-1$, the previous equation generates the $2D$
Ricci flow model (\ref{3.7}).

$Case2:k=0,$ $c_{3}=m+v$ and nonvanishing values for all the other constants 
$m,v,c_{1},c_{2}.$

The solution (\ref{3.18}) could take in this case the form:%
\begin{eqnarray}
A(x,y,t,u) &=&\left[ \frac{1}{2}e^{u}\xi ^{(-\frac{c}{m})}\right] \xi ^{(%
\frac{c}{m})}+\left[ \frac{1}{2}e^{u}\eta ^{(-\frac{c}{v})}\right] \eta ^{(%
\frac{c}{v})}=e^{u}  \nonumber \\
B(x,y,t,u) &=&\left[ \frac{1}{2}e^{u}\xi ^{(-\frac{c}{m})}\right] \xi ^{(%
\frac{c}{m})}+\left[ \frac{1}{2}e^{u}\eta ^{(-\frac{c}{v})}\right] \eta ^{(%
\frac{c}{v})}=\frac{dA}{du}=e^{u}  \label{3.24} \\
C(x,y,t,u) &=&D(x,y,t,u)=E(x,y,t,u)=F(x,y,t,u)=G(x,y,t,u)=0  \nonumber
\end{eqnarray}

In this second case, the following $2D$ evolution equation of type (\ref{2.1}%
) is generated:%
\begin{equation}
u_{t}=e^{u}u_{xy}+e^{u}u_{x}u_{y}=(e^{u}u_{x})_{y}  \label{3.25}
\end{equation}

\textbf{Remark 4: }The equations (\ref{3.23}) and (\ref{3.25}) generated by
the previous two cases correspond to the $2D$ nonlinear heat equation which
has the general expression \cite{NHE}:%
\begin{equation}
u_{t}=(g(u)u_{x})_{y}  \label{3.26}
\end{equation}%
It was proven \cite{NA} that the choices $g(u)=u^{n}$ and $g(u)=e^{u}$ are
the only two possible cases for which the Lie symmetries exist and more the
Lie operators have linear forms.

\subsection{The Lie symmetry problems for 2D convective-diffusion equation}

The second nonlinear application is represented by the $2D$
convective-diffusion equation \cite{cde}. It is a parabolic partial
differential equation, which describes physical phenomena where particles or
energy (or other physical quantities) are transferred inside a physical
system due to two processes: diffusion and convection. In the simpler case
when the diffusion coefficient is variable, the convection velocity is
constant and there are no sources or sinks, the equation takes the form:

\begin{equation}
u_{t}=uu_{2x}+uu_{2y}-vu_{x}  \label{4.1}
\end{equation}%
with diffusion coefficient $u$ and convective velocity $v=const.$ belongs to
the $Ox$ direction.

It is easy to remark that (\ref{4.1}) results from the general class of
equations (\ref{2.1}) by choosing the particular functions:%
\begin{eqnarray}
C(x,y,t,u) &=&D(x,y,t,u)=u,\text{ }F(x,y,t,u)=-v  \nonumber \\
A(x,y,t,u) &=&B(x,y,t,u)=E(x,y,t,u)=G(x,y,t,u)\equiv 0  \label{4.2}
\end{eqnarray}

\subsubsection{Lie symmetries for 2D convective-diffusion equation}

\bigskip In the conditions (\ref{4.2}) the general determining system (\ref%
{2.7}) for symmetries becomes:%
\begin{eqnarray}
\phi _{2u} &=&0  \nonumber \\
\xi _{y}+\eta _{x} &=&0  \nonumber \\
2u\xi _{x}-\phi &=&0  \nonumber \\
2u\eta _{y}-\phi &=&0  \label{4.3} \\
-\eta _{t}-v\eta _{x}+u\eta _{2x}+u\eta _{2y}-2u\phi _{yu} &=&0  \nonumber \\
-\xi _{t}-v\xi _{x}+u\xi _{2x}+u\xi _{2y}-2u\phi _{xu} &=&0  \nonumber \\
\phi _{t}+v\phi _{x}-u\phi _{2x}-u\phi _{2y} &=&0  \nonumber
\end{eqnarray}%
It has the solution:%
\begin{equation}
\xi =\frac{c_{1}}{2}(x-vt)+c_{2}y+c_{3},\text{ }\eta =\frac{c_{1}}{2}%
y-c_{2}(x-vt)+c_{4},\text{ }\phi =c_{1}u  \label{4.4}
\end{equation}%
In this case, the Lie symmetry generator takes the expression:%
\begin{equation}
U(x,y,t,u)=\frac{\partial }{\partial t}+\left( \frac{c_{1}}{2}%
(x-vt)+c_{2}y+c_{3}\right) \frac{\partial }{\partial x}+\left( \frac{c_{1}}{2%
}y-c_{2}(x-vt)+c_{4}\right) \frac{\partial }{\partial y}+c_{1}u\frac{%
\partial }{\partial u}  \label{4.5}
\end{equation}%
Consequently, the nonlinear convective-diffusion equation (\ref{4.1}) admits
the $4-$dimensional Lie algebra spanned by the operators shown below:%
\begin{eqnarray}
\,V_{1} &=&\left( \frac{x-vt}{2}\right) \frac{\partial }{\partial x}+\left( 
\frac{y}{2}\right) \frac{\partial }{\partial y}+u\frac{\partial }{\partial u}%
,  \label{4.6} \\
V_{2} &=&y\frac{\partial }{\partial x}-(x-vt)\frac{\partial }{\partial y}%
,\,V_{3}=\frac{\partial }{\partial x},V_{4}=\frac{\partial }{\partial y} 
\nonumber
\end{eqnarray}%
When the Lie algebra of these operators is computed, the only non-vanishing
relations are: 
\begin{equation}
\,\,[V_{3},V_{1}]=\frac{1}{2}V_{3},[V_{4},V_{1}]=V_{4},\text{ }%
[V_{2},V_{3}]=V_{4},\text{ }[V_{4},V_{2}]=V_{3}  \label{4.7}
\end{equation}

\subsubsection{Optimal system for convective-diffusion equation}

For this model the commutators of the symmetry operators (\ref{4.6}) are
given below in the Table 2:

\begin{tabular}{ccccc}
$\lbrack V_{i},V_{j}]$ & $V_{1}$ & $V_{2}$ & $V_{3}$ & $V_{4}$ \\ 
$V_{1}$ & $0$ & $0$ & $-V_{3}/2$ & $V_{4}$ \\ 
$V_{2}$ & $0$ & $0$ & $V_{4}$ & $-V_{3}$ \\ 
$V_{3}$ & $V_{3}/2$ & $-V_{4}$ & $0$ & $0$ \\ 
$V_{4}$ & $V_{4}$ & $V_{3}$ & $0$ & $0$%
\end{tabular}
\ \ \ \ \ \ Table 2: Lie brackets of the admitted symmetry algebra

Let us consider the linear combination of the symmetry generators:%
\begin{equation}
V=b_{1}V_{1}+b_{2}V_{2}+b_{3}V_{3}+b_{4}V_{4}  \label{6.1}
\end{equation}%
Our task is to simplify as many of the coefficients $b_{i}$ as possible
through judicious applications of adjoint maps to $V.$Suppose first that $%
b_{1}\neq 0$ in (\ref{6.1}). One may re-scale $b_{1}$ such that $b_{1}=1.$
We start with the combination:%
\begin{equation}
V^{(1)}=V_{1}+b_{2}V_{2}+b_{3}V_{3}+b_{4}V_{4}  \label{6.2}
\end{equation}%
If we act on $V^{(1)}$ by $Ad(\exp (2b_{3}V_{3})),$ we can make the
coefficient of $V_{3}$ vanish and we obtain the operator:%
\begin{equation}
V^{(2)}=V_{1}+b_{2}V_{2}+b_{4}^{\prime }V_{4},\text{ }b_{4}^{\prime
}=b_{4}+2b_{2}b_{3}  \label{6.3}
\end{equation}%
Using the adjoint representation (\ref{5.1}) for our model, no further
simplification is possible. Consequently, the $1-$dimensional subalgebra
spanned by $V$ with $b_{1}\neq 0$ is equivalent to the one spanned by $%
V_{1}+\alpha V_{2}+\beta V_{4},$ $\forall $ $\alpha ,\beta \in R.$

The remaining $1-$dimensional subalgebras are spanned by operators with $%
b_{1}=0$ which have the expressions:%
\begin{equation}
V^{(3)}=b_{2}V_{2}+b_{3}V_{3}+b_{4}V_{4}  \label{6.4}
\end{equation}%
Let us assume that $b_{2}\neq 0$ and scale to make $b_{2}=1.$ Now we act on $%
V^{(3)}$ by $Ad(\exp (b_{3}V_{4}))$ so that it is equivalent with the
operator:%
\begin{equation}
V^{(4)}=V_{2}+b_{4}V_{4}  \label{6.5}
\end{equation}%
Here, further simplification is possible. If we act on $V^{(4)}$ by $Ad(\exp
(-b_{4}V_{3}))$. In this case, we obtain the operator $V_{2}$ which is the
following $1-$dimensional subalgebra spanned $V$ with $b_{1}=0$ and $%
b_{2}=1. $

If we consider the case $b_{1}=b_{2}=0,$ $b_{3}\neq 0,b_{3}=1$, the
following generator is obtained:%
\begin{equation}
V^{(3)}=V_{3}+b_{4}V_{4}  \label{6.6}
\end{equation}%
If we act on $V^{(3)}$ by $Ad(\exp (\varepsilon V_{2})),$ where $\varepsilon 
$ is the solution of the equation%
\begin{equation}
\frac{b_{4}}{2}\varepsilon ^{2}+\varepsilon -b_{4}=0  \label{6.7}
\end{equation}%
we can vanish the coefficient of $V_{4}$ and we obtain the operator:%
\begin{equation}
V^{(4)}=b_{3}^{\prime }V_{3},\text{ }b_{3}^{\prime }=1+b_{4}\varepsilon -%
\frac{\varepsilon ^{2}}{2}  \label{6.8}
\end{equation}%
where $\varepsilon $ verified (\ref{6.7}).

Consequently, by setting $b_{1}=b_{2}=0,b_{3}=1$ in (\ref{6.1}) is generated 
$V_{3}$ which is the last subalgebra of the optimal system.

In conclusion, the optimal system of $1-$dimensional subalgebras for $2D$
convective-diffusion equation is:%
\begin{equation}
\{V_{2},V_{3},V_{1}+\alpha V_{2}+\beta V_{4},\forall \alpha ,\beta \in R\}
\label{6.9}
\end{equation}

\subsubsection{Invariant solutions for the convective-diffusion equation}

Through the reduced similarity method, each operator $\{V_{i},$ $i=\overline{%
1,4}\}$ can generate invariant solutions of the model. Let us illustrate for
our case what are the concrete forms of the similarity solutions generated
not by this base of operators, but by the set of the optimal $1D$
subalgebras (\ref{6.9}).

\begin{itemize}
\item Taking into account (\ref{4.6}), the symmetry operator $V_{2}$ has the
characteristic equations:%
\begin{equation}
\frac{dt}{0}=\frac{dx}{y}=\frac{dy}{vt-x}=\frac{du}{0}  \label{4.14}
\end{equation}%
In this second case, following the same procedure, we obtain also $3$
independent invariants with expressions:%
\begin{equation}
I_{1}=t,\text{ }I_{2}=vtx-\frac{x^{2}}{2}-\frac{y^{2}}{2},\text{ }I_{3}=u
\label{4.15}
\end{equation}%
With the notation $I_{2}\equiv z$ and $I_{3}=u\equiv g(t,z)$, the reduced
equation for $g(t,z)$ will take the form:%
\begin{equation}
g_{t}+(2z-v^{2}t^{2})gg_{2z}+2gg_{z}+v^{2}tg_{z}=0  \label{4.16}
\end{equation}%
It admits the solution:%
\begin{equation}
g(t,z)=\frac{2z-v^{2}t^{2}+2q_{1}}{4t+2q_{2}}  \label{4.17}
\end{equation}%
where $q_{1},q_{2},v$ are arbitrary constants.
\end{itemize}

Thereby, the second similarity solution corresponding to operator $V_{2}$
has the final form:

\begin{equation}
u(t,x,y)=\frac{2vtx-x^{2}-y^{2}-v^{2}t^{2}+2q_{1}}{4t+2q_{2}}  \label{4.18}
\end{equation}

\begin{itemize}
\item The operator $V_{3}$ from (\ref{4.6}) yields the characteristic
equations:%
\begin{equation}
\frac{dt}{0}=\frac{dx}{1}=\frac{dy}{0}=\frac{du}{0}  \label{4.191}
\end{equation}
\end{itemize}

Therefore also $3$ invariants are generated:%
\begin{equation}
I_{1}=t,\text{ }I_{2}=y,\text{ }I_{3}=u  \label{4.20}
\end{equation}%
Once again, expressing the last invariant $I_{3}$ as a function of the
others ones, we obtain the third similarity solution:%
\begin{equation}
u(t,y)=\frac{\frac{q_{1}}{2}y^{2}+q_{3}y+q_{4}}{q_{2}-q_{1}t}  \label{4.21}
\end{equation}%
with $q_{1},q_{2}$ arbitrary constants.

\begin{itemize}
\item Again on the basis of (\ref{4.6}), the last operator from (\ref{6.9}), 
$V_{1}+\alpha V_{2}+\beta V_{4},$ has the characteristic equations:%
\begin{equation}
\frac{dt}{0}=\frac{dx}{\alpha y+\frac{x-vt}{2}}=\frac{dy}{\frac{y}{2}+\alpha
(vt-x)+\beta }=\frac{du}{u}  \label{4.8}
\end{equation}%
By integrating these equations one obtains $3$ invariants with expressions:%
\begin{equation}
I_{1}=t,\text{ }I_{2}=\frac{\frac{y}{2}+\alpha (vt-x)+\beta }{\frac{x}{2}%
+\alpha y-\frac{vt}{2}},\text{ }I_{3}=\frac{u}{\left[ \frac{y^{2}}{2}+\alpha
(vt-x)+\beta \right] ^{2}}  \label{4.9}
\end{equation}%
By introducing the similarity variable $z\equiv $ $I_{2}$, designating the
invariant $I_{3}$ $=h(t,z)$ as a function of the other ones, the following
solution is obtained:%
\begin{equation}
u(t,x,y)=h(t,z)\left[ \frac{y^{2}}{2}+\alpha (vt-x)+\beta \right] ^{2}
\label{4.10}
\end{equation}%
Setting the derivatives of (\ref{4.10}) into the convective-diffusion
equation (\ref{4.1}), we obtain the following $(1+1)$ reduced equation for $%
h(t,z):$ 
\begin{equation}
h_{t}-2\left( \alpha ^{2}+\frac{1}{4}\right) z^{3}hh_{z}-4\left( \alpha ^{2}+%
\frac{1}{4}\right) zhh_{z}-2\left( \alpha ^{2}+\frac{1}{4}\right) h^{2}=0
\label{4.11}
\end{equation}%
The solution of the previous equation is:%
\begin{equation}
h(t,z)=\frac{-1}{2\left( \alpha ^{2}+\frac{1}{4}\right) t-\gamma }
\label{4.12}
\end{equation}%
with $\alpha ,\gamma $ arbitrary constants.

Using (\ref{4.10}), the invariant solution generated by the operator $%
V_{1}+\alpha V_{2}+\beta V_{4}$ is pointed out: 
\begin{equation}
u(t,x,y)=-\frac{1}{2\left( \alpha ^{2}+\frac{1}{4}\right) t-\gamma }\left[ 
\frac{y^{2}}{2}+\alpha (vt-x)+\beta \right] ^{2}  \label{4.13}
\end{equation}

where $\alpha ,\beta ,\gamma ,v$ arbitrary constants.
\end{itemize}

\subsubsection{Inverse symmetry problem for $2D$ convective-diffusion
equation}

Our aim is now to find the class of equations with generic form (\ref{2.1})
which admits the same symmetries with those corresponding to $2D$ nonlinear
convective-diffusion equation (\ref{4.1}). Consequently, we have to impose
that the coefficient functions (\ref{4.4}) which determine the base of
symmetry operators (\ref{4.6}) verify the general determining system (\ref%
{2.7}).

The solutions of differential system (\ref{2.7}) describe the coefficient
functions of the general evolutionary equation (\ref{2.1}) as follows:%
\begin{eqnarray}
A &=&B=0,\text{ }C=D=c_{3}u,  \nonumber \\
E(u) &=&\sqrt{u}\left[ c_{4}\cos \left( \frac{c_{2}}{c_{1}}\ln (u)\right)
-c_{5}\sin \left( \frac{c_{2}}{c_{1}}\ln (u)\right) \right]  \label{4.241} \\
F(u) &=&\sqrt{u}\left[ c_{4}\sin \left( \frac{c_{2}}{c_{1}}\ln (u)\right)
+c_{5}\cos \left( \frac{c_{2}}{c_{1}}\ln (u)\right) -v\right]  \nonumber \\
G(u) &=&c_{6}u  \nonumber
\end{eqnarray}%
where $c_{j},$ $j=\overline{1,6}$ and $v$ are arbitrary constants.

In particular, for $c_{3}=1,$ $c_{4}=c_{5}=c_{6}=0$ and arbitrary $c_{1}$
and $c_{2}$, the solution (\ref{4.241}) generates the $2D$ nonlinear
convective-diffusion equation (\ref{4.1}) discussed above.

\section{Conclusions}

This lecture intended to present some key aspects on how a dynamical systems
whose evolution is described by a nonlinear differential equation can be
studied using the symmetry method. The main steps which have to be done in
order to find a set of exact solutions are: (\textit{i}) determination of
the general form for the symmetry operator; (\textit{ii}) determination of
the optimal set of independent operators which can generates the minimal
subalgebras; (\textit{iii}) based on the optimal set of independent
operators and using the similarity reduction procedure, a complete set of
invariant solutions can be generated; (\textit{iv}) last but not least, a
special method can be applied in order to find the largest class of
nonlinear differential equations which belong to the same class as a given
equation in the sense of the symmetries they observe. This algorithm was
applied for two important examples of nonlinear $2D$ partial derivative
equations, the Ricci flow and the convective-diffusion equation. For the
first example the optimal system of subalgebras contains the same number of
generators, four, as the whole symmetry algebra. The optimal system of
symmetry subalgebra for the convective-diffusion equation has the dimension
three, despite the existence of four independent symmetry operators. In both
cases, the whole set of invariant solutions had been obtained.

\paragraph{Acknowledgements}

The authors are grateful for the financial support offered by the Romanian
Ministry of Education, Research and Innovation, through the National Council
for Scientific Research in Higher Education (CNCSIS), in the frame of the
Programme "Ideas", grant code ID 418/2008.

\bigskip\ 

\bigskip

\end{document}